\documentclass[onecolumn,preprintnumbers,amsmath,amssymb,eqsecnum,12pt]{revtex4}

\usepackage{epsf}
\usepackage{graphicx}  
\usepackage{dcolumn}   
\usepackage{bm}        

\begin{document}

\preprint{KU-TP 038}

\title{Black Holes without Mass and  Entropy in
Lovelock Gravity }

 \author{Rong-Gen Cai$^{a,b,}$\footnote{e-mail address:
cairg@itp.ac.cn}, Li-Ming Cao$^{b,}$\footnote{e-mail address:
caolm@phys.kindai.ac.jp},  Nobuyoshi Ohta$^{b,}$\footnote{e-mail
address: ohtan@phys.kindai.ac.jp}}


\address{$^{a}$
Key Laboratory of Frontiers in Theoretical
Physics, Institute of Theoretical Physics, Chinese Academy of
Sciences, P.O. Box 2735, Beijing 100190, China \\}

\address{$^{b}$
Department of Physics, Kinki University, Higashi-Osaka, Osaka
577-8502, Japan}

\vspace*{2.cm}
\begin{abstract}

We present a class of new black hole solutions in $D$-dimensional
Lovelock gravity theory. The solutions have a form of direct product
$\mathcal{M}^m \times \mathcal{H}^{n}$, where $D=m+n$,
$\mathcal{H}^n$ is a negative constant curvature space, and the
solutions are characterized by two integration constants.  When
$m=3$ and $4$, these solutions reduce to the exact black hole
solutions recently found by Maeda and Dadhich in Gauss-Bonnet
gravity theory. We study thermodynamics of these black hole
solutions. Although these black holes have a nonvanishing Hawking
temperature, surprisingly, the mass of these solutions always
vanishes. While the entropy also vanishes when $m$ is odd, it is a
constant determined by Euler characteristic of $(m-2)$-dimensional
cross section of black hole horizon when $m$ is even. We argue that
the constant in the entropy should be thrown away. Namely, when $m$
is even, the entropy of these black holes also should vanish. We
discuss the implications of these results.

\end{abstract}
\maketitle

\newpage
\section{Introduction}

With the development of string theory, supergravity and brane
world scenarios, over the past years, gravity theories have been
widely studied in higher dimensions. In the low-energy
approximation, Einstein general relativity naturally arises from
string theories. As corrections from massive states of string
theories and from loop expansions in string theories, some higher
derivative curvature terms also appear in the low-energy effective
action of string theories~\cite{Zwie, Lowenergylimit}. Therefore
it is of great interest to discuss potential roles of those higher
derivative terms in various aspects, for example, in black hole
physics and early universe. Indeed, there exist a lot of works on
higher derivative gravity theories in the literature.

In this paper we focus on a class of special higher derivative
gravity theory,  namely, Lovelock gravity~\cite{Love}, which is a
natural generalization of general relativity in higher dimensions
in the sense that the equations of motion of Lovelock gravity do
not contain more than second order derivatives with respect to
metric, as the case of general relativity.  The Lagrangian of
$D$-dimensional Lovelock gravity consists of the dimensionally
extended Euler densities
\begin{equation}
\label{1eq1} {\cal L} = \sum^p_{k=0}c_k {\cal L}_k,
\end{equation}
where $p\leq [(D-1)/2]$($[N]$ denotes the integral part of the
number $N$), $c_k$ are arbitrary constants with dimension of
$[\mathrm{Length}]^{2k-2}$, and ${\cal L}_k$ are the Euler densities
\begin{equation}
{\cal L}_k=\frac{1}{2^k}\sqrt{-g}\delta^{\mu_1\cdots
\mu_k\nu_1\cdots\nu_k}_{\lambda_1\cdots
\lambda_k\sigma_1\cdots\sigma_k}R^{\lambda_1\sigma_1}_{~~~~
\mu_1\nu_1}\cdots R^{\lambda_k\sigma_k}_{~~~~ \mu_k\nu_k}\, .
\label{Lk}
\end{equation}
Here, the generalized delta is totally antisymmetric in both sets of
indices. ${\cal L}_0=1$, so the constant $c_0$ is just the
cosmological constant. ${\cal L}_1$ gives us the usual curvature
scalar term, and for simplicity, we set $c_1=1$, while ${\cal L}_2$
is just the Gauss-Bonnet term. The Gauss-Bonnet term is argued to
appear in the low-energy action of heterotic string theory with a
positive coefficient~\cite{Zwie}. The equations of motion following
from the Lagrangian~(\ref{1eq1}) have the form
$\mathcal{G}_{\mu\nu}=0$, where
\begin{equation}
\mathcal{G}^{\mu}_{\nu}=\sum_{k=0}^{p}\frac{1}{2^{k+1}}
c_{k}\delta^{\mu\mu_1\cdots
\mu_k\nu_1\cdots\nu_k}_{\nu\lambda_1\cdots
\lambda_k\sigma_1\cdots\sigma_k}R^{\lambda_1\sigma_1}_{~~~~
\mu_1\nu_1}\cdots R^{\lambda_k\sigma_k}_{~~~~ \mu_k\nu_k} \,
.\label{em}
\end{equation}
Since the action of Lovelock gravity is the sum of the
dimensionally extended Euler densities, there are no more than
second order derivatives with respect to metric in its equations
of motion. Furthermore, the Lovelock gravity is shown to be free
of ghost when expanded on a flat space, evading any problems with
unitarity~\cite{Zwie,Des}.

Finding exact analytic solutions of any gravity theory is an issue
of long-standing interest. Indeed, there exist a lot of works to
discuss exact black hole solutions for Lovelock gravity in the
literature.  The static,  spherically symmetric black hole
solutions in the theory have been found
in~\cite{Des,Whee,Cai1,Cai4, Banados,CaiSoh,Zaneli} and discussed
\cite{Myers}, and topological nontrivial black holes have been
studied in~\cite{Cai1,Cai4,CaiSoh,Zaneli}. Some rotating solutions
in Gauss-Bonnet theory have been studied in~\cite{rotating,
Anabalon:2009kq}. However, a general rotating solution is still
absent even in Gauss-Bonnet gravity. See also \cite{others,
others1} for some other extensions including perturbative AdS
black hole solutions in the gravity theories with second order
curvature corrections. For a nice review of black holes in
Lovelock gravity, see~\cite{GG}.

Recently, Maeda and Dadhich presented a class of exact solutions in
Gauss-Bonnet gravity~\cite{Maeda:2006iw, Maeda:2006hj,
Dadhich:2007xf, Molina:2008kh} (Some similar black hole solutions
have been found in Codimension-2 brane world
theory~\cite{CuadrosMelgar:2007jx, CuadrosMelgar:2008kn}). They
assumed the spacetime has a direct product structure (one is a
four/three-dimensional spacetime, the other is a negative constant
curvature space), and then split the equations of motion into two
sets according to the direct product structure of spacetime. For a
suitable choice of those coefficients in Gauss-Bonnet gravity so
that the set of equations of motion for the four/three-dimensional
part is trivially satisfied, the set of the equations of motion for
the negative constant curvature space part then reduces to a single
equation. Solving the latter yields a class of new exact analytic
solutions for Gauss-Bonnet gravity. This class of the solutions has
two integration constants; one is argued to be related to the mass
of the solutions and the other behaves like a Maxwell charge. The
Maxwell charge is called ``Weyl charge" due to the existence of the
extra negative constant curvature space. This class of solutions is
quite different from the normal ones in the sense which will become
clear shortly. Quantum properties of those black hole solutions have
not yet been studied so far.

In this paper, we consider a general Lovelock theory instead of
the Gauss-Bonnet theory, and seek for  more general black hole
solutions and study their thermodynamics. The outline of the paper
is as follows. In Sec.~II, we present a class of new black hole
solutions in the general Lovelock theory,
following~\cite{Maeda:2006iw, Maeda:2006hj}. In Sec.~III, we study
thermodynamic properties of those black hole solutions.  Sec.~IV is
devoted to conclusion and discussion.

\section{General Black Hole Solutions}
\subsection{Equations of Motion}

Consider a $D(=m+n)$-dimensional spacetime $X$, locally homeomorphic
to $\mathcal{M}^m\times \mathcal{N}^n$. We assume the metric of this
spacetime has the form
\begin{equation}
\label{metric} ds^2=g_{ij}dx^i  dx^j+r_{0}^2 \gamma_{ab}dy^a dy^b\,
,
\end{equation}
where $g_{ij}dx^i  dx^j$ is the metric on $\mathcal{M}^m$ with the
coordinates $\{x^i,i=1,\cdots , m\}$, $r_{0}$ is a constant and
$\gamma_{ab}dy^a  dy^b$ is the metric on the $n$-dimensional space
$\mathcal{N}^n$ with the coordinates $\{y^a,a=,1,\cdots,n\}$.
$\mathcal{N}^n$ is a constant curvature space with curvature
$\bar{k}=\pm 1,0$. It is easy to find that the components of the
Riemann tensor for~(\ref{metric}) have the form
\begin{eqnarray}
\label{riemann1}
R_{ijkl}=\bar{R}_{ijkl},\quad R_{ij}{}{}^{kl}=\bar{R}_{ij}{}{}^{kl}\, ,\nonumber \\
R_{abcd}=\tilde{R}_{abcd},\quad
R_{ab}{}{}^{cd}=\tilde{R}_{ab}{}{}^{cd}\, ,
\end{eqnarray}
where $\bar{R}_{ijkl}$ and $\tilde{R}_{abcd}$ denote the components
of Riemann tensor on $\mathcal{M}^m$ and $\mathcal{N}^n$
respectively. For $\mathcal{N}^n$, we can write the
$\widetilde{R}_{abcd}$ as
\begin{equation}
\tilde{R}_{abcd}=\bar{k} r_{0}^2
\left(\gamma_{ac}\gamma_{bd}-\gamma_{ad}\gamma_{bc}\right), \quad
\tilde{R}^{ab}{}{}_{cd}= \frac{\bar{k}}{r_{0}^2} \delta^{ab}_{cd}\,
. \label{riemann2}
\end{equation}
According to the decomposition of the Riemann
tensor~(\ref{riemann1}), we can decompose the equations of
motion~(\ref{em}) into $m$-dimensional and $n$-dimensional parts:
\begin{eqnarray}
\mathcal{G}^{i}{}_{j}&\equiv&\sum_{k=0}^{p}\frac{1}{2^{k+1}}c_{k}\delta^{i
\mu_1\cdots \mu_k\nu_1\cdots\nu_k}_{j\lambda_1\cdots
\lambda_k\sigma_1\cdots\sigma_k}R^{\lambda_1\sigma_1}_{~~~~
\mu_1\nu_1}\cdots R^{\lambda_k\sigma_k}_{~~~~ \mu_k\nu_k}=0 \, , \nonumber \\
\mathcal{G}^{a}{}_{b}&\equiv&\sum_{k=0}^{p}\frac{1}{2^{k+1}}c_{k}\delta^{a
\mu_1\cdots \mu_k\nu_1\cdots\nu_k}_{b\lambda_1\cdots
\lambda_k\sigma_1\cdots\sigma_k}R^{\lambda_1\sigma_1}_{~~~~
\mu_1\nu_1}\cdots R^{\lambda_k\sigma_k}_{~~~~ \mu_k\nu_k}=0 \, .
\label{emsplit}
\end{eqnarray}
Other components (such as $\mathcal{G}^{i}{}_{a}$) automatically
vanish. Since $i,j$ run only in the range $\{1,\cdots ,
m\}$, and $R_{ijkl}$ can appear in the products ``$RR\cdots$" no
more than $q=[(m-1)/2]$ times, we have
\begin{eqnarray}
\label{2eq5}
\mathcal{G}^{i}{}_{j}&=&\sum_{k=0}^{p}\frac{c_{k}}{2^{k+1}}\delta^{i
\mu_1\cdots \mu_k\nu_1\cdots\nu_k}_{j\lambda_1\cdots
\lambda_k\sigma_1\cdots\sigma_k}R^{\lambda_1\sigma_1}_{~~~~
\mu_1\nu_1}\cdots R^{\lambda_k\sigma_k}_{~~~~ \mu_k\nu_k}\nonumber \\
&=&\sum_{t}^{q}\sum_{k=t}^{p} {k \choose t} \frac{c_{k}}{2^{k+1}}
\delta_{j m_{1} n_{1}\cdots m_{t} n_{t}  e_{1} \cdots e_{k-t}
f_{1}\cdots f_{k-t}}^{i k_1 l_1 \cdots k_t l_t  c_{1}\cdots c_{k-t}
d_{1} \cdots d_{k-t}}
R_{k_1 l_1}{}{}^{m_{1} n_{1}}\cdots R_{k_t l_t}{}{}^{m_{t} n_{t}}\nonumber \\
&& \times R_{c_{1}d_{1}}{}{}^{e_{1}f_{1}}\cdots
R_{c_{k-t}d_{k-t}}{}{}^{e_{k-t}f_{k-t}}\, .
\end{eqnarray}
Substituting Eqs.~(\ref{riemann1}) and (\ref{riemann2}) into (\ref{2eq5})
and using the identity
\begin{equation}
\delta_{\nu_{1}\cdots \nu_{p-1}\nu_{p}}^{\mu_{1}\cdots \mu_{p-1}
\mu_{p}}\delta_{\mu_{p-1}\mu_{p}}^{\nu_{p-1}\nu_{p}}
=2\left[r-(p-1)\right]\left[r-(p-2)\right]\delta_{\nu_{1}\cdots
\nu_{p-2}}^{\mu_{1}\cdots \mu_{p-2}}\, , \label{idenKD}
\end{equation}
where $r$ denotes the range of the index
($r=m$ for $\mathcal{M}^m$ and $r=n$ for $\mathcal{N}^n$), we have
\begin{eqnarray}
\label{emm}
\mathcal{G}^{i}{}_{j}&=&\sum_{t=0}^{q}\left[\sum_{k=t}^{p} {k
\choose t} c_k \frac{(D-m)!}{(D-m-2(k-t))!}
\left(\frac{\bar{k}}{r_{0}^2 }\right)^{k-t}
\right]\overline{\mathcal{G}}^{i}_{(t)j}\, ,
\end{eqnarray}
where
\begin{equation}
\overline{\mathcal{G}}^{i}_{(t)j}=\frac{1}{2^{t+1}}\delta^{ik_1 l_1
\cdots k_{t} l_{t}}_{jm_{1} n_{1}\cdots m_{t} n_{t}}R_{k_1
l_1}{}{}^{m_{1} n_{1}}\cdots R_{k_{t} l_{t}}{}{}^{m_{t} n_{t}}\, .
\end{equation}
Similarly, we have
\begin{eqnarray}
\mathcal{G}^{a}{}_{b}&=&\sum_{k=0}^{p}\frac{c_{k}}{2^{k+1}}\delta^{a
\mu_1\cdots \mu_k\nu_1\cdots\nu_k}_{b\lambda_1\cdots
\lambda_k\sigma_1\cdots\sigma_k}R^{\lambda_1\sigma_1}_{~~~~
\mu_1\nu_1}\cdots R^{\lambda_k\sigma_k}_{~~~~ \mu_k\nu_k}\nonumber \\
&=&\sum_{t=0}^{s}\sum_{k=t}^{p} {k \choose t} \frac{c_{k}}{2^{k+1}}
\delta_{b m_{1} n_{1}\cdots m_{t} n_{t} e_{1} \cdots e_{k-t}
f_{1}\cdots f_{k-t}}^{a k_1 l_1 \cdots k_t l_s  c_{1}\cdots c_{k-t}
d_{1} \cdots d_{k-t}}
R_{k_1 l_1}{}{}^{m_{1} n_{1}}\cdots R_{k_t l_t}{}{}^{m_{t} n_{t}}\nonumber \\
&&\times R_{c_{1}d_{1}}{}{}^{e_{1}f_{1}}\cdots
R_{c_{k-t}d_{k-t}}{}{}^{e_{k-t}f_{k-t}}\, ,
\end{eqnarray}
where $s=[m/2]$. Substituting Eqs.~(\ref{riemann1}),
(\ref{riemann2}) and (\ref{idenKD}) into the above equation,  we
can express it as
\begin{eqnarray}
\label{enn} \mathcal{G}^{a}{}_{b}&=& \frac{1}{2}\delta^{a}{}_{b}
\Bigg\{ \sum_{t=0}^{s}\left[\sum_{k=t}^{p} {k \choose t} c_k
\frac{(D-m-1)!}{(D-m-1-2(k-t))!} \left(\frac{\bar{k}}{r_{0}^2
}\right)^{k-t}\right]\bar{L}_t \Bigg\}\, ,
\end{eqnarray}
where
\begin{equation}
\bar{L}_t =\frac{1}{2^t}\delta^{k_1l_1\cdots
k_{t}l_{t}}_{m_1n_1\cdots m_{t}n_{t}}R_{k_1l_1}{}{}^{m_1n_1}\cdots
R_{k_{t}l_{t}}{}{}^{m_{t}n_{t}}\, . \label{Lbart}
\end{equation}
Note  that $\mathcal{G}^a_b$ is always proportional to
$\delta^a_b$, which is a crucial point to our discussions below.
Let us note that if the following equations are satisfied
\begin{equation}
\label{lineq}
0=A_t \equiv \sum_{k=t}^{p} {k \choose t} c_k
\frac{(D-m)!}{(D-m-2(k-t))!} \left(\frac{\bar{k}}{r_{0}^2
}\right)^{k-t},\quad t=0, \cdots, q\, ,
\end{equation}
then the equations of motion~(\ref{emm}) are always trivially satisfied.
These are $(q+1)$-linear equations for $c_0,\cdots,c_p $. Recall
$c_1=1$ and if we consider the case with
\begin{equation}
p=q+1=\left[\frac{m-1}{2}\right]+1\, ,
\end{equation}
then the equations (\ref{lineq}) indicate  that $c_k (k \neq 1)$ has
a unique expression in terms of  $\left(\bar{k}/r_{0}^2 \right)$ and
dimension $D$. When $\bar{k}=-1$ (which implies that $\mathcal{N}^n$
is a negative constant curvature space. It will be denoted by
$\mathcal{H}^n$ in the following discussion.) and $D\geq m+2$, we
find that all $c_k$ are positive. Some examples will be given soon.

Now we turn to the equations of $\mathcal{G}_{ab}$. Due to the
fact that $\mathcal{G}_{ab}$ is proportional to $\delta_{ab}$, the
equations $\mathcal{G}_{ab}=0$ reduce to a single equation
\begin{equation}
\label{2eq14}
 0=\sum_{i=0}^{s}\alpha_i \bar{L}_i\, ,
\end{equation}
where the coefficients $\alpha_i$'s are given by
\begin{equation}
\alpha_i=\sum_{k=i}^{p} {k \choose i} c_k
\frac{(D-m-1)!}{(D-m-1-2(k-i))!} \left(\frac{\bar{k}}{r_{0}^2
}\right)^{k-i}\, ,
\end{equation}
and $c_k$ are determined by solutions~(\ref{lineq}).

Let us further assume the $m$-dimensional metric $g_{ij}$ takes
the form
\begin{equation}
\label{m-2metric}
g=-f(r)dt^2+\frac{1}{f(r)}dr^2+r^2d\Sigma_{m-2}^2\, ,
\end{equation}
where $d\Sigma_{m-2}^2$ is the line element of $(m-2)$-dimensional
surface with constant scalar curvature $(m-2)(m-3)\delta$. Without
loss of generality, $\delta$ can be set to $\pm1$ or zero. It
is easy to find that the nonvanishing components of the Riemann
tensor are
\begin{equation}
R^{tr}{}{}_{tr}=-\frac{f^{''}}{2},\quad
R^{ti}{}{}_{tj}=R^{ri}{}{}_{rj}=-\frac{f^{'}}{2r}\delta_{j}^i,\quad
R^{ij}{}{}_{kl}=\frac{\delta-f}{r^2}\delta_{kl}^{ij}\, ,
\end{equation}
where the prime stands for derivative with respect to $r$.  The
Euler density then has the form
\begin{eqnarray}
\bar{L}_i&=&
\frac{(m-2)!}{(m-2-2i)!}\left(\frac{\delta-f}{r^2}\right)^{i}+4i
\frac{(m-2)!}{(m-1-2i)!} \left(-\frac{f^{'}}{2r}\right)
\left(\frac{\delta-f}{r^2}\right)^{i-1}
\nonumber \\
&+&2i \frac{(m-2)!}{(m-2i)!}
\left(-\frac{f^{''}}{2}\right)\left(\frac{\delta-f}{r^2}\right)^{i-1}
+4i(i-1)\frac{(m-2)!}{(m-2i)!}\left(-\frac{f^{'}}{2r}\right)^2
\left(\frac{\delta-f}{r^2}\right)^{i-2}\, .\nonumber\\
\end{eqnarray}
Defining
\begin{equation}
F(r)=\frac{\delta-f(r)}{r^2}\, ,
\end{equation}
 $\bar{L}_i$ can be rewritten as
\begin{equation}
\bar{L}_i = \frac{(m-2)!}{(m-2i)!}\frac{1}{r^{m-2}}\left(r^m
F(r)^i\right)^{''}\, .
\end{equation}
Finally Eq.~(\ref{2eq14}) becomes
\begin{equation}
\label{2eq21}
 0=\sum_{i=0}^{s}\widehat{\alpha}_{i}\left(r^m
F(r)^i\right)^{''}\, ,
\end{equation}
where
\begin{equation}
\widehat{\alpha}_{i}=\frac{(m-2)!}{(m-2i)!}\alpha_i\, .
\end{equation}
The solution to (\ref{2eq21}) is determined by the algebraic equation
\begin{equation}
\label{Fequ}
\sum_{i=0}^{s}\widehat{\alpha}_{i}F(r)^i=\frac{M}{r^{m-1}}+\frac{Q}{r^{m}}\,
,
\end{equation}
where $M, Q$ are two integration constants. Naively one may think
they are related to the mass and Weyl charge of the
solution~\cite{Maeda:2006iw,Dadhich:2000am, Shiromizu:1999wj},
respectively. But actually the integration constant $M$ has
nothing to do with the mass of the solution, which will be shown
shortly. The constant $Q$ may be positive, zero and negative. Here
some remarks are in order.

\indent(i) Since $\mathcal{G}^{a}_{b}\sim \delta^{a}_{b}$, the
equations  $\mathcal{G}^{a}_{b}=0$ reduce to a single equation
(\ref{2eq14}). This is very different from the normal case in
Lovelock gravity, where $\mathcal{G}^{a}_{b}$ is not
proportional to $\delta^{a}_{b}$ even under a spherical symmetric
assumption. For example, the equation $\mathcal{G}^{t}_{t}=0$ of
Lovelock theory will give a first order differential equation
like~\cite{Whee,Cai4}
\begin{equation}
0=\sum_{i=0}c_{i}\left(r^{m-1} F(r)^i\right)^{'}\, ,
\end{equation}
in the static, spherically symmetric case.  In that case, there is
only one integration constant, which is nothing but the mass of
the solution~\cite{Cai4}. In the present case, one has only
one traceless-like equation, which is a second order differential
equation. There is therefore one more integration constant $Q$ in
the present case.

\indent(ii) If $Q=0$, Eq.~(\ref{Fequ}) is very similar to the
corresponding one for static, spherically symmetric black hole
solutions in Lovelock theory~\cite{Whee,Cai4}. However, there are
two obvious differences: one is that here the coefficients
$\widehat{\alpha}_i$ are all fixed by $\bar{k}/r_0^2$ and $D$,
while in the normal case, those coefficients are free
parameters~\cite{Whee,Cai4}. We will show this below. The other is
that in Eq.~(\ref{Fequ}), the range of $i$ is $[0,\cdots,
s=[m/2]]$, while in the normal case, the range is $ [0,\cdots,
[(m-1)/2]]$. Therefore when $m$ is even, $i$ can take the value
$[m/2]$, which will not appear in the normal case.

\indent(iii) Since all $\widehat{\alpha}_0$ are fixed by
Eq.~(\ref{lineq}) and they are all positive constants, the
solutions are not asymptotically flat, but asymptotically AdS
spacetimes.


\subsection{ Black hole solution with $M=0$}

The spacetime (\ref{m-2metric}) describes a black hole provided
$f(r_+)=0$ and $f(r)>0$ with $r>r_+$. Here $r=r_+$ is called black
hole horizon. 
We can see from (\ref{Fequ}) that even if the ``mass" $M$ of the solution
vanishes, black hole horizon can still exist. To show this, let us discuss the solution
$F=F_0$ of Eq.~(\ref{Fequ}) with $M=0$:
\begin{equation}
\sum_{i=0}^{s}\widehat{\alpha}_{i}F^i_0=\frac{Q}{r^{m}}
\end{equation}
with horizons. Assume $f_0(\bar{r})=0$ at some positive $\bar{r}$,
and we have
\begin{equation}
\bar{r}^2F_0(\bar{r})= \delta \quad \mathrm{or}\quad
F_0(\bar{r})=\frac{\delta}{\bar{r}^2}\, .
\end{equation}
That is,  $\bar{r}$ must satisfy the following equation
\begin{equation}
\label{2eq27}
\bar{r}^{m}\sum_{i=0}^{s}\widehat{\alpha}_{i}\left(\frac{\delta}{\bar{r}^2}\right)^i=Q\,
.
\end{equation}
In order for the equation to hold, the constant $Q$ must satisfy
some constraints. We will discuss these constraints in the cases
of $\delta=0$ and $\delta=\pm1$, respectively.

\noindent (i).  $\delta=0$. This case is simple. In this case, only one
term in (\ref{2eq27}) remains. We have
\begin{equation}
\bar{r}^{m}\widehat{\alpha}_{0}=Q\, ,
\end{equation}
Since $\widehat{\alpha}_{0}>0$, this indicates a positive
$\bar{r}$ exists provided $Q>0$, and $\bar{r}$ is just the
horizon radius $r_+$.

\noindent (ii).  $\delta=\pm 1$. If $m$ is odd, one then has
$s=(m-1)/2$, and
\begin{equation}
\label{2eq29}
\bar{r}^{m}\widehat{\alpha}_{0}\pm\bar{r}^{m-2}\widehat{\alpha}_{1}+\cdots
+(\pm1)^s\bar{r}\widehat{\alpha}_{s}=Q\, .
\end{equation}
Obviously, because all coefficients $\{\widehat{\alpha}_{0},
\widehat{\alpha}_{1}\cdots \widehat{\alpha}_{s}\}$ are positive,
and especially  $\widehat{\alpha}_{0}>0$, Eq.~(\ref{2eq29}) has
at least one positive root $\bar{r}$ if  $Q>0$.
The black hole horizon $r_+$ is just the largest positive root
of Eq.~(\ref{2eq29}).

On the other hand, if $m$ is even, one then has $s=m/2$, and
\begin{equation}
\bar{r}^{m}\widehat{\alpha}_{0}\pm\bar{r}^{m-2}\widehat{\alpha}_{1}+\cdots
+(\pm1)^{s-1}\bar{r}^2\widehat{\alpha}_{s-1}+(\pm1)^s\widehat{\alpha}_{s}=Q\,
.
\end{equation}
From the theory of polynomial, the above equation has at least one
negative root and one positive root if
$\widehat{\alpha}_{0}\left[(\pm1)^s\widehat{\alpha}_{s}-Q\right]<0$.
Recall $\widehat{\alpha}_{0}>0$, and this condition can be always
satisfied if $Q>\widehat{\alpha}_{s}$. Namely, black hole horizon
exists in this case.

In summary, black hole horizon always exists provided
$Q-\widehat{\alpha}_{s}>0$, even when the parameter $M$ vanishes.


\subsection{Examples of exact solutions}

To be more explicit, in this subsection, we give some simple examples
of exact solutions given in (\ref{Fequ}).

\subsubsection{The case of $m=3$, $p=2$, $D\geq 5$}

In this case, Eqs.~(\ref{lineq}) give
\begin{equation}
\label{c0c2r0D} c_0=\frac{1}{2}(D^2-3D-6)r_0^{-2}\, ,\quad
c_2=\frac{1}{2(D-3)(D-4)}r_0^2\, .
\end{equation}
We then have
\begin{equation}
\widehat{\alpha}_0=\frac{2(2D-9)}{3(D-3)}r_0^{-2}\, ,\quad
\widehat{\alpha}_1=\frac{2}{D-3}\, .
\end{equation}
Equation~(\ref{Fequ}) for $F$ becomes
\begin{equation}
\widehat{\alpha}_0+\widehat{\alpha}_1
F=\frac{M}{r^2}+\frac{Q}{r^3}\, ,
\end{equation}
which has the solution
\begin{equation}
f=\frac{1}{\widehat{\alpha}_1}\left(-M-\frac{Q}{r}+\widehat{\alpha}_0
r^2\right)\, .
\end{equation}
Here we have used the fact that  the $(m-2)$-dimensional constant
curvature space is always Ricci flat for $m=3$, i.e., $\delta=0$.
Since $\widehat{\alpha}_0$ is always positive, this solution is
just a BTZ black hole deformed by the additional charge $Q$. This
kind of solution has been obtained in~\cite{Maeda:2006hj}.

\subsubsection{The case of $m=4$, $p=2$, $D\geq 6$}

In this case, Eq.~(\ref{lineq}) give
\begin{equation}
\label{c0c2r0Dm4} c_0=\frac{1}{2}(D^2-5D-2)r_{0}^{-2},\quad
c_2=\frac{1}{2(D-4)(D-5)}r_{0}^{2}\, .
\end{equation}
We then have
\begin{equation}
\widehat{\alpha}_0=\frac{2D-11}{3(D-4)}r_{0}^{-2}\, ,\quad
\widehat{\alpha}_1=\frac{2}{D-4}\, ,\quad
\widehat{\alpha}_2=\frac{1}{(D-4)(D-5)}r_{0}^{2}\, .
\end{equation}
The equation for $F$ becomes
\begin{equation}
\widehat{\alpha}_0+\widehat{\alpha}_1
F+\widehat{\alpha}_2F^2=\frac{M}{r^3}+\frac{Q}{r^4}\, ,
\end{equation}
which has the solution
\begin{equation}
F(r)=-\frac{\widehat{\alpha}_1}{2\widehat{\alpha}_2}
\left(~1\mp\sqrt{1-\frac{4\widehat{\alpha}_0
\widehat{\alpha}_2}{\widehat{\alpha}_1^2}+\frac{4
\widehat{\alpha}_2M}{\widehat{\alpha}_1^2~r^3}+\frac{4
\widehat{\alpha}_2Q}{\widehat{\alpha}_1^2~r^4}}~\right)\, ,
\end{equation}
\begin{equation}
\label{m4solution} f(r)=\delta+\frac{\widehat{\alpha}_1
r^2}{2\widehat{\alpha}_2}\left(~1\mp\sqrt{1-\frac{4\widehat{\alpha}_0
\widehat{\alpha}_2}{\widehat{\alpha}_1^2}+\frac{4
\widehat{\alpha}_2M}{\widehat{\alpha}_1^2~r^3}+\frac{4
\widehat{\alpha}_2Q}{\widehat{\alpha}_1^2~r^4}}~\right)\, .
\end{equation}
This solution with two branches is just the one recently obtained
by Maeda and Dadhich in~\cite{Maeda:2006iw}.  It is easy to see
that for $D\ge 6$, we have
\begin{equation}
\widehat{\alpha}_1^2-4\widehat{\alpha}_0\widehat{\alpha}_2=\frac{4}{3(D-4)(D-5)}>0\,
,
\end{equation}
So the vacuum AdS solution ($M=Q=0$) always exists.

\subsubsection{The case of $m=5$, $p=3$, $D\geq 7$}

Equations~(\ref{lineq}) in this case lead to
\begin{eqnarray}
c_0&=&\frac{D^4-10D^3+11D^2+22D+360}{3(D^2-7D+4)}r_{0}^{-2}\, ,\nonumber \\
c_2&=&\frac{1}{(D^2-7D+4)}r_{0}^{2}\, ,\nonumber\\
c_3&=&\frac{1}{3(D^2-7D+4)(D-5)(D-6)}r_{0}^{4}\, .
\end{eqnarray}
Then the corresponding $\widehat{\alpha}$ are
\begin{eqnarray}
\widehat{\alpha}_0&=&\frac{4(5D^2-67D+225)}{5(D^2-7D+4)(D-5)}r_{0}^{-2}\, ,\nonumber \\
\widehat{\alpha}_1&=&\frac{8(2D-13)}{(D^2-7D+4)(D-5)}\, ,\nonumber \\
\widehat{\alpha}_2&=&\frac{12}{(D^2-7D+4)(D-5)}r_{0}^{2}\, .
\end{eqnarray}
The equation for $F$ is still in second order and has the solution
\begin{equation}
F(r)=-\frac{\widehat{\alpha}_1}{2\widehat{\alpha}_2}
\left(~1\mp\sqrt{1-\frac{4\widehat{\alpha}_0
\widehat{\alpha}_2}{\widehat{\alpha}_1^2}+\frac{4
\widehat{\alpha}_2M}{\widehat{\alpha}_1^2~r^4}+\frac{4
\widehat{\alpha}_2Q}{\widehat{\alpha}_1^2~r^5}}~\right)\, ,
\end{equation}
\begin{equation}
f(r)=\delta+\frac{\widehat{\alpha}_1
r^2}{2\widehat{\alpha}_2}\left(~1\mp\sqrt{1-\frac{4\widehat{\alpha}_0
\widehat{\alpha}_2}{\widehat{\alpha}_1^2}+\frac{4
\widehat{\alpha}_2M}{\widehat{\alpha}_1^2~r^4}+\frac{4
\widehat{\alpha}_2Q}{\widehat{\alpha}_1^2~r^5}}~\right)\, .
\end{equation}
This solution is an example with the third order Lovelock term.
Note that when $m=5$, the solution also has two branches.  In
addition, it is also easy to see that the vacuum AdS solution
exists, because
$$
\widehat{\alpha}_1-4\widehat{\alpha}_0\widehat{\alpha}_1
=\frac{64 (5D-34)}{5(D^2-7D+4)(D-5)}>0
$$
for $D\ge 7$.

\section{Thermodynamic properties of black hole solutions}

\subsection{Naive consideration: $m$-dimensional black holes}

The black hole spacetime has a direct product form $\mathcal{M}^m
\times \mathcal{H}^{n}$, where $\mathcal{H}^{n}$ is a negative
constant curvature space with a constant radius $r_0$. From the
point of view of usual Kaluza-Klein dimensional reduction, the
thermodynamics for the whole spacetime is equivalent to that for
$m$-dimensional black hole with redefined gravitational constant
$G_m=G_{m+n}/{\rm Vol}(n)$. Here $G_m$ and $G_{m+n}$ are
gravitational constants in $m$ dimensions and $(m+n)$ dimensions,
respectively, while ${\rm Vol}(n)$ is the volume of the constant
curvature space $\mathcal{H}^{n}$. In this subsection, we will
discuss the black hole thermodynamics from the point of view of $m$
dimensions.

Assume that black hole has a horizon at $r_+$, which is the
largest positive root of $f(r)=0$. The horizon radius then must
satisfy
\begin{equation}
F(r_{+})=\frac{\delta}{r_{+}^2}\, , \quad \mathrm{or} \quad
r_{+}^2F(r_{+})= \delta\, .
\end{equation}
The Hawking temperature of the black hole can be easily calculated
by Euclidean method. To avoid conical singularity at the horizon,
the period of Euclidean time should be $ \beta = 4\pi/f'(r_+) $,
and the Hawking temperature is just the inverse of the period.
This way we get the temperature of the black hole
\begin{equation}
\label{temperature}
\mathfrak{T}=\frac{1}{\beta}=\frac{1}{4\pi}f^{'}(r_{+})=-\frac{1}{4\pi}
\left(\frac{2\delta}{r_{+}}+r_{+}^2F^{'}(r_{+})\right)\,
.
\end{equation}
To get the explicit form of the Hawking temperature in terms of black hole horizon,
we have to give the
expression of $F^{'}$. Taking derivative on both sides of Eq.
(\ref{Fequ}) with respect to $r$, one has
\begin{equation}
F^{'}(r_{+})=-\frac{(m-1)M r_{+} + m Q}{r_{+}^{m+1}\sum_{i=1}^{s}
i~\widehat{\alpha}_{i}\left(\frac{\delta}{r_{+}^2}\right)^{i-1}}\, .
\end{equation}
On the other hand, from the equation (\ref{Fequ}), we have
\begin{equation}
\label{massparameter} M=-\frac{Q}{r_{+}}+
r_{+}^{m-1}\sum_{i=0}^{s}\widehat{\alpha}_{i}\left(\frac{\delta}{r_{+}^2}\right)^i\,
.
\end{equation}
Thus we can express $F^{'}(r_+)$ as
\begin{equation}
F^{'}(r_{+})=-\frac{Q+(m-1)r_{+}^m\sum_{i=0}^{s}
\widehat{\alpha}_{i}\left(\frac{\delta}{r_{+}^2}\right)^{i}}{r_{+}^{m+1}\sum_{i=1}^{s}
i~\widehat{\alpha}_{i}\left(\frac{\delta}{r_{+}^2}\right)^{i-1}}\, ,
\end{equation}
and the Hawking temperature has the form
\begin{equation}
\mathfrak{T}=\frac{1}{4\pi r_{+}\sum_{i=1}^{s}
i~\widehat{\alpha}_{i}\left(\frac{\delta}{r_{+}^2}\right)^{i-1}}
\left(\sum_{i=0}^{s}(m-2i-1)
\widehat{\alpha}_{i}\delta\left(\frac{\delta}{r_{+}^2}\right)^{i-1}
+\frac{Q}{r_{+}^{m-2}}\right)\,
.
\end{equation}
Clearly, if we choose a suitable $Q$, the Hawking temperature may
vanish.  This case corresponds to the ``extremal" black holes with
vanishing Hawking temperature. For example, when $m=4$, choosing
\begin{equation}
Q=-3\widehat{\alpha}_0 r_{+}^4-\widehat{\alpha}_1
r_{+}^2\delta+\widehat{\alpha}_2\delta^2
\end{equation}
will lead to a vanishing temperature.

To get the mass of the black hole, we expand the metric $g_{00}$
in the large $r$ limit, subtract the corresponding one for a suitable
reference background solution $F_b$, and then read off the
mass with the coefficient in front of some power of the radial
coordinate $r$. Here we choose the vacuum AdS solutions with
vanishing $M$ and $Q$ as the reference background, i.e.,
\begin{equation}
\sum_{i=0}^{s}\widehat{\alpha}_i F_b^i = 0 \, .
\end{equation}
For large $r$, we can expand $F$ as $F=F_b+\Delta F$ with the
leading order correction $\Delta F$. We arrive at
\begin{equation}
\Delta F\sum_{i=0}^{s} i ~\widehat{\alpha}_iF_b^{i-1}= \Delta
F\widehat{\alpha}=\frac{M}{r^{m-1}}   \, .
\end{equation}
Here the constant $\widehat{\alpha}$ is given by
\begin{equation}
\widehat{\alpha}=\sum_{i=0}^{s} i ~\widehat{\alpha}_iF_b^{i-1}\, .
\end{equation}
For solutions in some branch, $\widehat{\alpha}$ may be negative.
However, we only consider the cases with positive
$\widehat{\alpha}$ here. So we find the expansion of metric around
the background as
\begin{equation}
g_{tt}-g_{(b)tt}= -f+f_{b}=r^2\Delta F \approx
\frac{M}{\widehat{\alpha}~ r^{m-3}}=\frac{16\pi
G_{m}\mathfrak{M}}{(m-2)V_{m-2}r^{m-3}}\, .
\end{equation}
Thus we find that the mass of the black hole $\mathfrak{M}$ has a
relation to $M$ as
\begin{equation}
\label{massm} \frac{M}{\widehat{\alpha}}=\frac{16\pi
G_{m}\mathfrak{M}}{(m-2)V_{m-2}}\, ,
\end{equation}
where $V_{m-2}$ is the volume of $(m-2)$-dimensional cross section
of horizon surface. The mass can be expressed  in terms of horizon
radius $r_+$ and $Q$ as
\begin{equation}
\mathfrak{M}=\frac{(m-2)V_{m-2}}{16\pi
G_{m}\widehat{\alpha}}\left(-\frac{Q}{r_{+}}+
r_{+}^{m-1}\sum_{i=0}^{s}\widehat{\alpha}_{i}\left(\frac{\delta}{r_{+}^2}\right)^i\right)\,
,
\end{equation}
and its variation with respect to the horizon radius is
\begin{equation}
\left(\frac{\partial \mathfrak{M}}{\partial
r_{+}}\right)_{Q}=\frac{(m-2)V_{m-2}}{16\pi
G_{m}\widehat{\alpha}}r_{+}^{m-4}\left(\sum_{i=0}^{s}(m-2i-1)
\widehat{\alpha}_{i}\delta\left(\frac{\delta}{r_{+}^2}\right)^{i-1}+\frac{Q}{r_{+}^{m-2}}\right)\,
.
\end{equation}
Since we are dealing with black holes in higher derivative gravity
theory, the well-known area formula for black hole entropy breaks down.
Let us try to obtain the black hole entropy by integrating the first
law of the black hole thermodynamics
\begin{equation}
\label{integration}
\mathfrak{S}=\int \mathfrak{T}^{-1} d
\mathfrak{M} =\int^{r_+} \mathfrak{T}^{-1}\left(\frac{\partial
\mathfrak{M}}{\partial r_{+}}\right)_{Q}dr_{+}\, .
\end{equation}

\noindent(i). When $m$ is even, $s$ takes the value $m/2$. The
integral gives
\begin{eqnarray}
\label{3eq16}
\mathfrak{S}&=&\int \mathfrak{T}^{-1} d \mathfrak{M}
=\int^{r_+} \mathfrak{T}^{-1}\left(\frac{\partial
\mathfrak{M}}{\partial
r_{+}}\right)_{Q}dr_{+}\nonumber\\
&=&\frac{V_{m-2}}{4
G_{m}\widehat{\alpha}}\left[\sum_{i=1}^{s-1}\frac{m-2}{m-2i}
i~\widehat{\alpha}_{i}\delta ^{i-1}r_+^{m-2i} +s(s-1)
\widehat{\alpha}_s~\delta^{s-1}~\mathrm{ln}
\left(r_+^2\right)\right] +\mathfrak{S}_0\, .
\end{eqnarray}
The last term $\mathfrak{S}_0$ is an integration constant. Note
that here a logarithmic term appears, which comes from the fact that
$s$ can take the value $m/2$.

\noindent (ii). When $m$ is odd, $s$ is $(m-1)/2$. In this case,
the integral gives
\begin{eqnarray}
\label{3eq17}
 \mathfrak{S}&=&\int \mathfrak{T}^{-1} d \mathfrak{M}
=\int^{r_+} \mathfrak{T}^{-1}\left(\frac{\partial
\mathfrak{M}}{\partial
r_{+}}\right)_{Q}dr_{+}\nonumber\\
&=&\frac{V_{m-2}}{4
G_{m}\widehat{\alpha}}\left[\sum_{i=1}^{s}\frac{m-2}{m-2i}
i~\widehat{\alpha}_{i}\delta ^{i-1}r_+^{m-2i}\right]+ \mathfrak{S}_0
\, .
\end{eqnarray}
Here $\mathfrak{S}_0$ is also an integration constant. Note that
here the integration constant $\mathfrak{S}_0$ should  be set to zero
because when the black hole horizon shrinks to zero,
the entropy of the black hole should vanish~\cite{Cai1}. However,
the integration constant cannot be fixed by the same argument in
the case of even $m$, due to the existence of the logarithmic term
in the black hole entropy. In addition, let us notice that when
the black hole horizon is a Ricci flat surface, namely,
$\delta=0$, not only does the logarithmic term disappear in
(\ref{3eq16}), but also both (\ref{3eq16}) and (\ref{3eq17}) give
an entropy proportional to horizon area. This is also a
general feature of black hole entropy in Lovelock
gravity~\cite{Cai4,Cai1}.

The entropy expressions (\ref{3eq16}) and (\ref{3eq17}) look quite
similar to the entropy formula of static, spherically symmetric
black holes in Lovelock gravity~\cite{Cai4}, except for the
logarithmic term in (\ref{3eq16}). The appearance of the
logarithmic term is strange, although such a term appears in the
entropy expressions of black holes in Horava-Lifshitz gravity
theory~\cite{CCO}, while the latter is not a full diffeomorphism
invariant theory.  For a diffeomorphism invariant gravity theory,
Wald showed that black hole entropy is a Noether
charge~\cite{Wald:1993nt}; further a well-known entropy formula
was developed~\cite{Wald:1993nt, Iyer:1994ys}. By Wald's entropy
formula, black hole entropy is given by some integral on the black
hole horizon. Therefore black hole entropy must be a function of
horizon geometry and a logarithmic term will never appear in
Wald's entropy formula. This may cause suspicion whether the
results given above are valid or not.

Let us notice that the above way to obtain black hole solution in
$m$ dimensions is quite different from the usual Kaluza-Klein
dimensional reduction.
In the usual Kaluza-Klein theory, with the assumption of
direct product of the manifold $\mathcal{M}^m\times \mathcal{H}^n$,
one gets reduced action by integrating the total action over the extra
space $\mathcal{H}^n$. Certainly here too, with the assumption of
direct product structure of the spacetime, one can get reduced
action from the total action (\ref{1eq1}). This reduced action is an
$m$-dimensional version of Lovelock gravity as given below in Eq.~(\ref{reducedaction}).
We can get equations of motion (\ref{emm}) only for $m$-dimensional part
but not the $n$-dimensional part by variation of this reduced action.
Obviously these are nothing but the usual equations of motion of the Lovelock
gravity in $m$ dimensions with special coefficients $A_{t}$'s. For example, when
$m=4$, these equations are the Einstein equations with cosmological constant
\begin{equation}
\label{eins}
A_0 g_{ij} + A_1 E_{ij}=0\, ,
\end{equation}
where $E_{ij}$ is the Einstein tensor in four dimensions, and $A_0$
and $A_1$ are given in (\ref{lineq}). Because our solutions are
obtained for $A_0=A_1=0$, Eqs.~(\ref{eins}) have no information on
our solutions. Therefore the field equations for $m$-dimensional
part is trivially satisfied, while the nontrivial solutions come
from the trace equation of gravitational field for the
$n$-dimensional part, which is not obtained from the reduced action.
So the reduced action certainly exists, but it does not give
solutions in this paper. If we naively omit the extra dimensions in
our solutions, the corresponding $m$-dimensional local
diffeomorphism-invariant ``effective action" is absent, and we have
to consider the whole $(m+n)$-dimensional theory. This is very
different from the usual Kaluza-Klein theory in which the effective
action is just the reduced action.
As a result, we cannot
simply use Wald's entropy formula to get the entropy of the $m$-dimensional
black hole. On the other hand, Lovelock theory is
diffeomorphism invariant and Wald's entropy formula is applicable in
the whole $(m+n)$ dimensions. In the following subsections, we discuss
thermodynamics of the black holes by Euclidean action and
Wald's entropy formula in $(m+n)$ dimensions, and find quite different
and surprising results.

\subsection{Entropy of $(m+n)$-dimensional black holes}

From the viewpoint of the whole $D(=m+n)$ dimensions, to study
thermodynamics of these black holes is straightforward. The
temperature of the black holes is the same as the one in
Eq.~(\ref{temperature}) because it is determined by the horizon
geometry only. In Lovelock gravity, Wald's entropy formula can be
expressed as~\cite{Jacobson:1993xs}
\begin{equation}
\label{entropywald} \mathfrak{S}=\sum_{k=1}^{p}4\pi k c_k \int
d^{D-2}x\mathcal{L}_{k-1}(\tilde{h})\, ,
\end{equation}
where $\mathcal{L}_{k-1}(\tilde{h})$ has the same form as
(\ref{Lk}) except that metric is replaced by $\tilde{h}$, which is
the induced metric on the $(D-2)$-dimensional cross section of the
horizon.  The induced metric $\tilde h$ is
\begin{equation}
\tilde{h}= \tilde{q}_{ij}dz^idz^j + r_{0}^2 \gamma_{ab}dy^a dy^b\, ,
\end{equation}
where $\tilde{q}_{ij}$ is the induced metric of the cross section of
the horizon in the $m$-dimensional part.

By the similar procedure to get (\ref{emm}) and (\ref{enn}), we have
\begin{eqnarray}
\mathfrak{S}&=&\sum_{t=0}^{w}\sum_{k=t+1}^{p}\Bigg{\{}{k-1 \choose
t}4\pi k c_k
\frac{(D-m)!}{(D-m-2(k-1-t))!}\left(\frac{\bar{k}}{r_0^2}\right)^{k-t-1}\nonumber\\
&~~~~~~~\times& \int d^{D-m}yr_0^{D-m} \sqrt{\gamma}\int
d^{m-2}z\mathcal{L}_{t}(\tilde{q})\Bigg{\}}\, ,
\end{eqnarray}
where $w=[(m-2)/2]$. Define $\ell=t+1$, we have
\begin{eqnarray}
\label{ent}
\mathfrak{S}&=&\sum_{\ell=1}^{w+1}\Bigg{\{}4\pi
\ell\Bigg{[}\sum_{k=\ell}^{p}{k \choose \ell} c_k
\frac{(D-m)!}{(D-m-2(k-\ell))!}\left(\frac{\bar{k}}{r_0^2}\right)^{k-\ell}
\Bigg{]}\nonumber\\
&~~~~~~~\times& \int d^{D-m}yr_0^{D-m} \sqrt{\gamma}\int
d^{m-2}z\mathcal{L}_{\ell-1}(\tilde{q})\Bigg{\}}\, .
\end{eqnarray}
Thus the entropy can be expressed as
\begin{eqnarray}
\label{3eq22}
\mathfrak{S}=\sum_{\ell=1}^{w+1}\Bigg{\{}4\pi \ell
A_{\ell} \int d^{D-m}yr_0^{D-m} \sqrt{\gamma}\int
d^{m-2}z\mathcal{L}_{\ell-1}(\tilde{q})\Bigg{\}}\, ,
\end{eqnarray}
where $A_{\ell}$ is defined in Eq.~(\ref{lineq}), from which we have
\begin{equation}
\label{Alequation} A_{\ell}=0\, ,\quad \ell=1,\cdots ,[(m-1)/2]\, .
\end{equation}
Thus we find from the entropy (\ref{3eq22}) that when $m$ is odd,
$\mathfrak{S}=0$, a vanishing entropy!

When $m$ is even, the entropy is
\begin{equation}
\label{entropy1} \mathfrak{S}=2\pi m
\left(r_0^{D-m}\Omega_{D-m}\right)c_{m/2}\chi(\Sigma_{m-2})\, .
\end{equation}
where $\Omega_{D-m}$ is the volume of $\mathcal{H}^n$, and
$\chi(\Sigma_{m-2})$ is the integration of Euler characteristic on
the $(m-2)$-dimensional cross section of horizon surface, i.e.,
\begin{equation}
\chi(\Sigma_{m-2})=\int_{\Sigma_{m-2}}
d^{m-2}z\mathcal{L}_{(m/2-1)}(\tilde{q})\, .
\end{equation}
For the cross section of horizon surface $\Sigma_{m-2}$, which is a
constant curvature space, $\chi(\Sigma_{m-2})$ is constant, while in
the case of $\Sigma_{m-2}$ being a closed manifold, $\Sigma_{m-2}$
need not be a constant curvature space, and in that case,
$\chi(\Sigma_{m-2})$ is the Euler number of $\Sigma_{m-2}$ up to a
constant factor. For example, when $m=4$, $D=6$, we have
\begin{equation}
\mathfrak{S}= 64\pi^2\cdot \delta\cdot c_2 \cdot(r_0^2 \Omega_2)\, .
\end{equation}
Clearly we see that when $\Sigma_2$ is a Ricci flat space, i.e.
$\delta =0$, the constant entropy vanishes. Here the constant
means that it is independent of the horizon radius $r_+$ and
charge $Q$.

Now we argue that the constant entropy is meaningless
for black hole thermodynamics and should be dropped. One simple
reason is that when the cross section of horizon surface
$\sigma_{m-2}$ is a negative constant curvature space, $\chi$ is
negative, giving a negative entropy which does not make sense in
thermodynamics. Another reason is provided by the following
example. Consider a four-dimensional Schwarzschild black hole
solution. In this case, the Euler density is the Gauss-Bonnet
term. If one considers the contribution of the Gauss-Bonnet term
to the black hole entropy, besides the usual area entropy, one has
an additional constant $\mathfrak{S}= 64\pi^2 c_2$ from
(\ref{entropy1}), where $c_2$ is the Gauss-Bonnet coefficient.
Such a constant term remains even when the black hole horizon goes
to zero. Both of these clearly indicate that the constant entropy from the
horizon topology should be dropped when black hole thermodynamics
is concerned.

Let us illustrate these discussions by two examples of $m=3$ and 4.

\subsubsection{The case of $m=3$, $p=2$, $D\ge 5$}

In this case, the metric $\tilde{h}$ is very simple, which is just
the metric of the constant curvature space $\mathcal{H}^n$ plus the
metric of a circle, i.e.,
\begin{equation}
\tilde{h}= dz^2 + r_{0}^2 \gamma_{ab}dy^a dy^b\, ,
\end{equation}
and Wald's entropy (\ref{entropywald}) becomes
\begin{equation}
\label{m34entropy} \mathfrak{S}=4\pi  \int d^{D-2}x+8\pi c_2\int
d^{D-2}xR(\tilde{h})\, .
\end{equation}
Here we have set $c_1$ to unity as before, and $R(\tilde{h})$ is
the scalar curvature of the metric $\tilde{h}$. It is easy to find
\begin{equation}
R(\tilde{h})=(D-3)(D-4)\left(\frac{\bar{k}}{r_0^2}\right)\, ,
\end{equation}
thus we have
\begin{equation}
\mathfrak{S}=4\pi
(r_0^{D-3}\Omega_{D-3})V_1\left[1+2c_2(D-3)(D-4)\left(\frac{\bar{k}}{r_0^2}\right)\right]
\, ,
\end{equation}
where $V_1$ is the volume of the circle. Considering $\bar{k}=-1$,
and the explicit relation among $c_2$, $r_0$ and $D$ in Eq.~(\ref{c0c2r0D}),
the entropy identically vanishes.

 Actually, for
$m=3$, we have $w=0$, so $\ell$ in Eq.~(\ref{3eq22}) can take the
value $1$ only. Since $A_{1}=0$ by Eq.~(\ref{Alequation}), we reach
a vanishing entropy from (\ref{3eq22}), as we have just shown above.

\subsubsection{The case of $m=4$, $p=2$, $D\ge 6$}

In this case, the metric $\tilde{h}$ consists of the metric of the
constant curvature space $\mathcal{H}^n$ and the metric of a
2-dimensional constant curvature space $\Sigma_2$, i.e.,
\begin{equation}
\tilde{h}= \tilde{q}_{ij}dz^idz^j + r_{0}^2 \gamma_{ab}dy^a dy^b\, ,
\end{equation}
and Wald's entropy is the same as (\ref{m34entropy}). Now, the
scalar curvature $R(\tilde{h})$ becomes
\begin{equation}
R(\tilde{h})= R(\tilde{q}) +
(D-4)(D-5)\left(\frac{\bar{k}}{r_0^2}\right)\, ,
\end{equation}
where $R(\tilde{q})$ is the scalar curvature of the metric
$\tilde{q}$. Thus the entropy~(\ref{ent}) has the form
\begin{equation}
\mathfrak{S}=4\pi
(r_0^{D-4}\Omega_{D-4})V_2\left[1+2c_2(D-4)(D-5)\left(\frac{\bar{k}}{r_0^2}\right)\right]+
8\pi c_2 (r_0^{D-4}\Omega_{D-4}) \int
\sqrt{\tilde{q}}d^2zR(\tilde{q}) \, .
\end{equation}
Using the explicit relation among $c_2$, $r_0$ and $D$ in equation
(\ref{c0c2r0Dm4}), and $\bar{k}=-1$, we see that only the last term
remains
\begin{equation}
\label{entropym4} \mathfrak{S}= 8\pi c_2 (r_0^{D-4}\Omega_{D-4})
\int \sqrt{\tilde{q}}d^2zR(\tilde{q}) \, .
\end{equation}
In other words, in this case, the entropy is totally determined by
the integration of the Euler characteristic $R(\tilde{q})$ on
$\Sigma_2$, in agreement with (\ref{entropy1}).  Note that the the
two-dimensional induced horizon $\tilde{q}$ is a constant curvature
space with scalar curvature $2\delta$. The entropy can be further
expressed as
\begin{equation}
\mathfrak{S}= 64\pi^2 \cdot \delta \cdot c_2 \cdot
(r_0^{D-4}\Omega_{D-4})\, .
\end{equation}

In fact, in the case of $m=4$, one has  $w=1$, and $\ell$ in
(\ref{3eq22}) can take values $1$ and $2$. We have $A_1=0$ from
(\ref{Alequation}), while $A_2=c_2$. By the general entropy
expression (\ref{3eq22}), we arrive at the same result as
Eq.~(\ref{entropym4}). As argued above, the constant entropy does
not make sense in black hole thermodynamics,  we should drop it and
conclude that the physical entropy is zero.

\subsection{Mass and Euclidean action of $m+n$ dimensional black holes}

In this subsection we show another surprising result that the mass
of these black holes also vanishes. To do this, we employ the
Euclidean approach to black hole thermodynamics.  The Euclidean
action $I_E$ of the black holes includes two parts, the bulk and
boundary parts,
\begin{equation}
I_E=I+B\, ,
\end{equation}
where $I$ is the bulk action, while $B$ denotes the boundary term.
The bulk part is given by
\begin{equation}
\label{reducedaction}
I=-\left(r_0^{D-m}\Omega_{D-m}\right)
\left\{\sum_{t=0}^{s}A_t\int\sqrt{g}d^{m}x\bar{L}_{t}\right\}\, ,
\end{equation}
where $\bar{L}_{t}$ is given by (\ref{Lbart}). With the metric
(\ref{m-2metric}), we get
\begin{equation}
\label{bulkaction} I=-\left(r_0^{D-m}\Omega_{D-m}V_{m-2}\right)
\left\{\sum_{t=0}^{s}A_t\frac{(m-2)!}{(m-2t)!}\int d\tau  dr
\left(r^m F^t\right)''\right\}\, ,
\end{equation}
where $V_{m-2}$ is the volume of the $\Sigma_{m-2}$ with unit
radius. In general,  the boundary term $B$ is a little bit
complicated. For simplicity,  we only consider here the case that
the highest derivative term is the Gauss-Bonnet term, i.e., we
deal with the cases with $m=3$ and $m=4$. In that case, the
boundary term is given by~\cite{Myers:1987yn,
Davis:2002gnGravanis:2002wy}
\begin{equation}
B=-2\int_{\partial X}d^{D-1}v\sqrt{h}\left[K+2c_2
\left(J-2E_{\mu\nu}K^{\mu\nu}\right)\right]\, ,
\end{equation}
where we have set $c_1=1$ and  $h$ is the induced metric on a
timelike boundary $\partial X$. The tensor $K^{\mu\nu}$ is the
extrinsic curvature of the boundary, and $K$ is its trace.
We denote by $J$ the trace of the tensor
\begin{equation}
J_{\mu\nu}=\frac{1}{3}\left(2K
K_{\mu\lambda}K^{\lambda}{}_{\nu}+K_{\lambda\sigma}K^{\lambda\sigma}
K_{\mu\nu}-2K_{\mu\lambda}K^{\lambda\sigma}K_{\sigma\nu}-K^2
K_{\mu\nu}\right)\, .
\end{equation}
Tensor $E_{\mu\nu}$ is the Einstein tensor of the induced metric
$h$. With this boundary term, the variation principle is well
defined for the Gauss-Bonnet gravity.

In the following calculations, we consider a boundary $\partial X$
with a given $r \gg r_+$, and take the limit of $r \to \infty$ at
the end of calculations.

\subsubsection{The case of $m=3$, $p=2$, $D\ge 5$}

The case of $m=3$ is quite simple. Both the bulk and boundary terms
identically vanish $I=B=0$.  In fact, in this case all $A_{t}$ are
zero and therefore the bulk term (\ref{bulkaction}) vanishes. For
the boundary term $B$, after some calculation, it is not hard to
find the tensor $J_{\mu\nu}=0$, and
\begin{equation}
K=\frac{2f+rf'}{2r\sqrt{f}}\, ,\nonumber
\end{equation}
\begin{equation}
2c_2(J-2E_{\mu\nu}K^{\mu\nu})=-c_2(D-3)(D-4)\left(\frac{2f+rf'}{rr_0^2\sqrt{f}}\right)\,
.
\end{equation}
So the boundary term is given by
\begin{equation}
B=- \beta
r_0^{D-3}\Omega_{D-3}V_1(2f+rf')\left[1-\frac{2c_2(D-3)(D-4)}{r_0^2}\right]\,
,
\end{equation}
where $\beta$ is the period of Euclidean time. Again, considering
the relation among $c_2$, $r_0$ and $D$ in equation (\ref{c0c2r0D}),
the boundary term has no contribution to the Euclidean action. Thus
the Euclidean action of the black hole solutions is always zero,
which leads to the conclusion that the energy and entropy of the
black holes always vanish.

In fact, when $m$ is odd, since all $A_t=0$, both the bulk and
boundary terms always vanish. Thus the result with vanishing energy
and entropy is universal for odd $m$.

\subsubsection{The case of $m=4$, $p=2$, $D\ge 6$}

When $m=4$, the bulk action (\ref{bulkaction}) reduces to
\begin{equation}
I= 4 c_2 \left(r_0^{D-4}\Omega_{D-4}V_2\right)\beta
\left[f'(\delta-f)\right]_{r=\infty} - 4 c_2
\left(r_0^{D-4}\Omega_{D-4}V_2\right)\beta \left[f'(\delta
-f)\right]_{r=r_+}\, ,
\end{equation}
where $\beta$ is the period of Euclidean time as before.  It is easy
to find the trace of the extrinsic curvature is given by
\begin{equation}
K=\frac{4f+rf'}{2r\sqrt{f}}\, ,
\end{equation}
and after some calculations, we can obtain
\begin{eqnarray}
2c_2(J-2E_{\mu\nu}K^{\mu\nu})&=&
\frac{c_2}{r^3r_0^2\sqrt{f}}\bigg{\{}rf'\left[2\delta r_0^2-(D-4)(D-5)r^2\right]\nonumber\\
&&-2f \left[2(D-4)(D-5)r^2-4r_0^2+rr_0^2f' \right]\bigg{\}}\, .
\end{eqnarray}
Therefore, the boundary term $B$ is
\begin{eqnarray}
B&=&-2 \left(r_0^{D-4}\Omega_{D-4}V_2\right)\beta \Bigg{[}\left(1
-\frac{2c_2(D-4)(D-5)}{r_0^2}\right)\left(2rf+\frac{1}{2}r^2f'\right)\nonumber\\
&&+2c_2 f'(\delta-f)+\frac{8c_2}{r}f\Bigg{]}_{r=\infty}\, .
\end{eqnarray}
Note that the equation (\ref{c0c2r0Dm4}) or $A_1=0$ gives
$1-2c_2(D-4)(D-5)/r_0^2=0$. Thus the boundary term reduces to
\begin{equation}
B=-4c_2 \left(r_0^{D-4}\Omega_{D-4}V_2\right)\beta \left[
f'(\delta-f)+\frac{4}{r}f\right]_{r=\infty}\, .
\end{equation}
We thus get the total action
\begin{equation}
I_E=I+B=-4c_2 \left(r_0^{D-4}\Omega_{D-4}V_2\right)\beta \left[
f'(\delta -f)\right]_{r=r_+}-\left.16\beta c_2
\left(r_0^{D-4}\Omega_{D-4}V_2\right)
\frac{f}{r}\right|_{r=\infty}\, .
\end{equation}
For our solutions, the second term in the right hand side is
divergent when $r \to \infty $. This divergence can be removed by
the background subtraction method. By subtracting the contribution
from the reference background with vanishing $M$ and $Q$, we find
that the second term does not make any contribution to the Euclidean
action and only the first term remains.

Note that  $\beta=4\pi/f'(r_+)$ and $f=0$ at the horizon. The first
term can be expressed as
\begin{equation}
\label{3eq36}
 I_E= -64\pi^2\cdot \delta\cdot c_2 \cdot\left(r_0^{D-4}\Omega_{D-4}\right)\, .
\end{equation}
This is a constant independent of temperature. Considering the
relation between the Euclidean action $I_E$ and free energy $F$:
$I_E=\beta F$, we immediately see that the energy of the black holes
always vanishes, while the Euclidean action (\ref{3eq36}) gives the
constant entropy (\ref{entropy1}) found by Wald's formula.

Thus by calculating Euclidean action and Wald's entropy, we have
shown that mass and entropy of these black hole solutions presented
in the previous section  vanish identically.


\section{Conclusion and Discussion}

In this paper we have presented a class of black hole solutions in
($m+n$)-dimensional Lovelock gravity.  The black hole solutions have
a direct product structure $\mathcal{M}^m \times \mathcal{H}^{n}$,
where $\mathcal{H}^n$ is a negative constant curvature space with a
constant radius. When $m=3$ and $4$, these solutions reduce to those
recently found by Maeda and Dadhich~\cite{Maeda:2006iw,
Maeda:2006hj, Dadhich:2007xf, Molina:2008kh} in Gauss-Bonnet
gravity. We have obtained these black hole solutions in a way as
follows. We first decompose the equations of motion into two sets,
one for $m$-dimensional part and the other for $n$-dimensional part.
Then imposing constraints on the coefficients of higher curvature
terms in Lovelock gravity so that the set of equations of motion for
the $m$-dimensional part is trivially satisfied, we solve the trace
equation for the $n$-dimensional part and obtain the black hole
solutions.  Since the trace equation is a second order differential
equation, integrating the equation gives rise to two integration
constants $M$ and $Q$.

We have tried to understand the physical meaning of the two
integration constants by studying thermodynamics of these black hole
solutions. The black holes we have found are exact solutions in
$(m+n)$-dimensional Lovelock gravity theory. Naively considering the
solution without the extra dimensions, it appeared that the mass
were proportional to parameter $M$ as in Eq.~(\ref{massm}). By using
the first law, we then found that the entropy of the black hole
would have logarithmic term when $m$ is even. However, we have
argued that this naive result is not valid for our solutions because
the equations of motion for the $m$-dimensional part are trivially
satisfied and our solutions come from the trace equation for the
$n$-dimensional part. As a result the effective action for the
$m$-dimensional part does not make any sense for the black hole
solutions. We should consider the black hole solutions in the point
of view of the whole $(m+n)$-dimensional spacetime.  It is not
surprising because our theory is intrinsically $(m+n)$-dimensional.
We cannot naively neglect the extra dimensions as in the case of the
usual Kaluza-Klein theory in which the thermodynamics in higher
dimensions and lower dimensions are equivalent.

Then by employing Euclidean action
approach to black hole thermodynamics and Wald's entropy formula,
we have found an astonishing result that both mass and entropy of these black
holes always vanish identically although there exists a nonvanishing Hawking
temperature for these black holes. Here it may be worth mentioning
that when $m$ is even, by Wald's entropy formula, the black hole
has a constant entropy coming from the topological structure of the black
hole horizon. But we have argued that the constant entropy should be
neglected from the point of view of black hole thermodynamics.

Black hole solutions with a nonvanishing temperature and always
vanishing mass and entropy look strange. But such a situation has
happened in a class of Lifshitz black holes in $R^2$
gravity~\cite{Cai:2009ac}. There these authors got the solution by
adjusting the coupling constant of $R^2$ term to a critical value.
Note that the same thing happens here in the class of Lovelock
black hole solutions since we have chosen a special set of coupling
coefficients of higher derivative terms in order to get our solutions.

Let us now try to understand such a phenomenon that a black hole has
a nonvanishing temperature, but vanishing mass and entropy. Recall
that in the $R^2$ gravity considered in \cite{Cai:2009ac}, the
resulting Lifshitz black hole solution satisfied $1+2\alpha R=0$,
where $\alpha$ is the coefficient of the term $R^2$, which is a
crucial point to give the zero entropy of the black hole. Let us
notice that in the $R^2$ theory, the factor $1+2\alpha R$ is nothing
but the effective coupling constant for some polarized graviton.
From the equations of motion, it is easy to see that the effective
gravitational constant turns to be $G_{eff}= G/(1+2\alpha R)$. As a
result, $1+2\alpha R=0$ implies that the effective gravitational
constant is divergent for the class of solutions with $1+2\alpha
R=0$. Wald's entropy is equal to a quarter of the horizon area in
units of the effective gravitational coupling~\cite{BGH}. This is
the reason why the entropy of the Lifshitz black hole has a
vanishing entropy. Furthermore, because of $G_{eff} \to \infty$,
from the point of view of background fluctuations, kinetic terms of
those fluctuations always vanish, and only potential terms remain.
This indicates that there is no dynamics for those fluctuations. In
other words, there are no excitations of the background spacetime.
This might be an interpretation why the black hole has no entropy.
While the Hawking temperature (surface gravity) of a black hole is
purely determined by black hole geometry in the sense that the
Hawking temperature is just the inverse period of the Euclidean time
of the black hole, the first law of thermodynamics enforces that a
black hole has a vanishing mass (energy) if its entropy is zero.

Let us turn to the black hole spacetime discussed in the present
paper. In fact the same happens here. The part of
$\mathcal{H}^{n}$ is a trivial negative constant curvature space.
The effective gravitational field equations for the part of
$m$-dimensional black hole spacetime are trivially satisfied in
the sense that those coefficients in front of some gravitational
tensors like Einstein tensor are identically zero [see (\ref{emm})
and (\ref{lineq})]. To see this more clearly, one may refer to the
simple case with Gauss-Bonnet gravity discussed in
\cite{Maeda:2006iw}. The vanishing coefficients are correspondent
to the factor $1+2\alpha R$ discussed above for $R^2$ gravity.
Therefore, due to the special reduction used to find the black
hole solutions in the present paper, these effective coupling
constants from the $m$-dimensional point of view identically
vanish. In this sense, the effective gravitational constant
$G_{eff}$ diverges as in the case of the $R^2$ gravity. Then the
same story goes on as the case of the $R^2$ gravity and these
black holes have vanishing entropy and mass.

If our arguments are true, our above discussions and those in
Ref.~\cite{Cai:2009ac} both have important consequence on our
understanding of the microscopic degrees of freedom of black hole
entropy. According to 't Hooft's brick wall model~\cite{tHooft},
black hole entropy might come from statistical degrees of freedom of
quantum fluctuations outside the black hole, namely if there is no
such degrees of freedom of quantum fluctuations, there is no
contribution to the entropy. The black hole entropy is not merely
determined by the geometry of the horizon.  In the examples
discussed above, the Bekenstein-Hawking geometry entropy of the
black holes always vanishes and the zero entropy is found to be
closely related to the fact that the effective gravitational
coupling constants are infinity such that any fluctuations are
forbidden, there are totally no physical degrees of freedom
associated with quantum fluctuations. Thus our results provide
evidence that black hole entropy comes from statistical degrees of
freedom of quantum fluctuations around the black hole. No doubt, it
is worthwhile to further investigate this interesting issue.

\section*{Acknowledgements}
RGC is supported partially by grants from NSFC, China (No.
10535060, No. 10821504 and No. 10975168) and a grant from MSTC,
China (No. 2010CB833004). LMC and NO were supported in part by
the Grants-in-Aid for Scientific Research Fund of the JSPS
Nos. 20540283 and 21$\cdot$\,09225, and also by the Japan-U.K.
Research Cooperative Program. This work was finished during RGC's
visit to Kinki University with a JSPS invitation fund.

\end{document}